\documentclass[aps,twocolumn,amsfonts,amsmath,amssymb,superscriptaddress,10pt,floatfix]{revtex4-1}
\usepackage{times}
\usepackage[english]{babel}
\usepackage[utf8]{inputenc}
\usepackage{amsmath}
\usepackage{amsfonts}
\usepackage{amssymb}
\usepackage{amsthm}
\usepackage{graphics}
\usepackage[dvips]{graphicx}
\usepackage{color}
\usepackage{rotating}
\usepackage[normalem]{ulem}
\usepackage {ulem}
\usepackage{import}
\begin{document}
\title{
Effect of entropy on the nucleation of cavitation bubbles in water under tension
}
\author{Georg Menzl}
\affiliation{Faculty of Physics and Center for Computational Materials Science, University of Vienna, Boltzmanngasse 5, 1090 Vienna, Austria}
\author{Christoph Dellago}
\thanks{To whom correspondence should be addressed:\\ christoph.dellago@univie.ac.at}
\affiliation{Faculty of Physics and Center for Computational Materials Science, University of Vienna, Boltzmanngasse 5, 1090 Vienna, Austria}
\begin{abstract}
Water can exist in a metastable liquid state under tension for long times before the system relaxes into the vapor via cavitation, i.e., bubble nucleation.
Microscopic information on the cavitation process can be extracted from experimental data by use of the nucleation theorem,
which relates measured cavitation rates to the size of the critical bubble.
To apply the nucleation theorem to experiments performed along an isochoric path, for instance, in cavitation experiments in mineral inclusions, knowledge of the bubble entropy is required.
Using computer simulations, we compute the entropy of bubbles in water 
as a function of their volume
over a wide range of tensions from free energy calculations.
We find that the bubble entropy is an important contribution to the free energy which significantly lowers the barrier to bubble nucleation, thereby facilitating cavitation.
Furthermore, the bubble entropy per surface area depends on the curvature of the liquid--vapor interface, decreasing approximately linearly with its mean curvature over the studied range of bubble volumes.
At room temperature, the entropy of a flat liquid--vapor interface at ambient pressure is very similar to that of critical bubbles over a wide range of tensions, which justifies the use of the former as an approximation when interpreting data from experiments.
Based on our simulation results, we obtain an estimate for the volume of the critical bubble from experimentally measured cavitation rates.
\end{abstract}
\maketitle
\section{Introduction}
Due to its highly cohesive nature, water can sustain strongly negative pressures exceeding $-120\, {\rm MPa}$ \cite{GreenAngellScience1990,ZhengAngellScience1991,AlvarengaBodnarJCP1993,AzouziCaupinNaturePhys2013,MercuryShmulovichNATO2014,PallaresCaupinPNAS2014} for long times before the system eventually decays into the vapor phase via cavitation, i.e., the nucleation of small vapor bubbles.
In the initial stages of this process, the formation of bubbles is opposed by a free energy barrier, and only when this barrier is crossed can the bubble grow to macroscopic size. The height of this barrier, which is due to the free energetic cost of the bubble--liquid interface, determines the cavitation rate.
As a result of the rapid nature of cavitation and the small size of the critical bubble at physically relevant conditions, the microscopic mechanism of bubble nucleation in water cannot be directly observed in experiments. However, some microscopic insights can be gained from the measured cavitation rates by use of the nucleation theorem,\cite{KashchievJCP1982} which links the change in the height of the free energy barrier in response to a change in external conditions to properties of the critical bubble, i.e., the bubble corresponding to the top of the free energy barrier. 
When experiments are performed at constant temperature, e.g., in acoustic cavitation setups where regions of negative pressure, $p$, are created by sound waves \cite{CaupinHerbertCRPhys2006,DavittBalibarJCP2010}, the nucleation theorem relates the pressure derivative of the barrier height $g^\ast$ to the volume $v^\ast$ of the critical bubble,
\begin{equation}\label{eq:nuctheorem_isothermal}
  \frac{\mathrm{d} g^\ast}{\mathrm{d} p} 
  = v^\ast.
\end{equation}
Thus, an estimate for the volume of the critical bubble can be readily obtained from the measured cavitation rates, which depend exponentially on the barrier heights.
To date the most precise and reliable experiments on cavitation in water have been carried out using water in mineral inclusions
\cite{GreenAngellScience1990,ZhengAngellScience1991,AlvarengaBodnarJCP1993,AzouziCaupinNaturePhys2013,MercuryShmulovichNATO2014}.
In these experiments, cavities containing liquid water and a vapor bubble are heated until the bubble disappears and the inclusion is filled by homogeneous fluid. On subsequent cooling, which essentially follows an isochoric path since the volume of the cavities barely changes over the temperature range, the liquid is stretched until cavitation occurs. When this happens, the tension on the liquid is released resulting in a phase separated system (bubble in liquid).
Along the isochoric cooling path, the nucleation theorem relates the properties of the critical bubble to the temperature derivative of the barrier height \cite{KashchievJCP2006},
\begin{align}\label{eq:nuctheorem_isochoric}
  \frac{\mathrm{d} g^\ast}{\mathrm{d} T} 
  &= v^\ast \left(\frac{\partial p}{\partial T}\right)_V - s^\ast,
\end{align}
where 
$s^\ast = s(v^\ast)$ is the entropy of the critical bubble and $( \partial p/\partial T ) \vert_V$ is the temperature derivative of the pressure of the metastable liquid at constant volume.
Hence, in order to infer the volume $v^\ast$ of the critical bubble from the measured cavitation rates, an estimate for the entropy $s^\ast$ of the critical bubble is required (an estimate for $( \partial p/\partial T ) \vert_V$ can be obtained from an equation of state \cite{DavittBalibarJCP2010,PallaresCaupinPCCP2016}). Since $s^\ast$ cannot be directly measured in experiment, computer simulations are a suitable choice to obtain the information needed for the interpretation of experimental data.
In this work, we perform computer simulations to obtain the entropy $s(v)$ as a function of the bubble volume $v$ over a wide range of tensions. 
Our simulations show that, at ambient temperature, the effect of the bubble entropy on the free energetics of nanoscale bubbles is comparable in magnitude to the enthalpic contributions.
Since the enthalpic contributions favor the metastable liquid whereas the entropy drives the system towards the vapor, the free energy barrier impeding nucleation is a direct result of these competing effects.
Furthermore, we find that the bubble entropy 
per unit area decreases linearly with the mean curvature $1/r$ of the bubble, suggesting a connection to the number of free OH groups at the vapor--liquid interface. 
Bubble entropies determined in our simulations suggest that the entropy of the flat liquid--vapor interface can be used as an estimate 
in the interpretation of experimental data, as done in Ref.~\onlinecite{AzouziCaupinNaturePhys2013}, thereby providing the basis for the application of the nucleation theorem to inclusion experiments.
\section{Methods}\label{sec:numericalmethods}
 
In this section, we first provide details on the computational techniques used to obtain our results, followed by a definition of the bubble volume employed in our simulations, which is calibrated to correspond to the true volume of bubbles on the molecular level. To conclude, we describe how the entropy of a flat vapor--liquid interface at ambient conditions, which serves as a point of comparison to the computed bubble entropies, is obtained.
\subsection{Simulation details}
We simulated a system of $N = 2000$ water molecules in the {\it NpT} ensemble using the rigid, non-polarisable TIP4P/2005 model~\cite{AbascalVegaJCP2005}, which has been shown to predict many properties of liquid water accurately, in particular the liquid--vapor surface tension and enthalpy~\cite{VegaMiguelJCP2007}.
The free energy of cavitation as a function of the volume of the largest bubble was computed by umbrella sampling~\cite{TorrieValleauJCP1977} with a hybrid Monte Carlo (HMC) scheme~\cite{DuaneKennedyPhysLettB1987,MehligHeermannPRB1992}.
During each HMC move, the center of mass and angular velocities of each molecule were drawn from the Maxwell--Boltzmann distribution corresponding to the desired temperature of $T = 296.4 \, {\rm K}$ and the system was propagated according to Newton's equations of motion using a time-reversible quaternion-based integrator~\cite{Kamberaj2005,Miller2002,Martyna1996,Omelyan2002} that maintains the rigid geometry of water molecules.
Each HMC step consisted of three molecular dynamics integration steps with a time step $\delta t = 7 \, {\rm fs}$ and pressure was kept constant by isotropic volume moves accepted according to the Metropolis criterion \cite{MetropolisTellerJCP1953}.
Sampling was enhanced by replica exchange moves \cite{GeyerThompsonJAmAstat1995} between neighboring windows and 
histograms obtained for the individual windows were pieced together using a self consistent histogram 
method.\cite{FerrenbergSwendsenPRL1989}
\subsection{Order parameter}\label{subsec:op}
In order to study homogeneous bubble nucleation in water, we use the volume of the largest bubble as a local order parameter. A committor analysis performed at ambient temperature and negative pressures suggests that the volume of the largest bubble constitutes a good reaction coordinate for cavitation\cite{ournucleationpaper}. 
Estimates for the volume $v$ of every bubble present in the system are obtained by use of the V-method,\cite{ourfirstpaper} which is designed to give thermodynamically consistent bubble volumes, i.e., volumes consistent with the nucleation theorem.
The V-method consists of two steps. First, the preliminary volume $\xi$ of the largest bubble in the system is assessed by using a grid-based order parameter similar to the procedure employed in Ref.~\onlinecite{WangFrenkelJPhysChemB2009}.\footnote{Note that we altered the nomenclature to facilitate readability: $v$ and $\xi$ in this work correspond to $V^{\rm V}_{\rm bubble}$ and $v$ in Ref.~\onlinecite{ourfirstpaper}, respectively.}
To this end, we superimpose a three-dimensional grid consisting of $52^3$ points onto the system
and determine vapor-like points, i.e., points where no liquid-like molecules are within a cutoff radius of $3.35\,$\AA.
(A molecule is considered to be liquid-like if it has at least one molecule within the same cutoff radius.)
These vapor-like grid points are then clustered and each of the resulting clusters constitutes a bubble with volume equal to the total volume of the grid cells belonging to the respective cluster.
The largest one of these volumes is the preliminary volume $\xi$ which we use to track the progress of the nucleation process.
Note that a number of suitable bubble detection procedures have been suggested in the literature \cite{AbascalValerianiJCP2013,torabi13a,torabi13b,GonzalezBresmeJCP2015}, which can all be used to obtain the preliminary volume $\xi$ of the largest bubble in the system. 
In the second step, we calibrate the bubble volume $v$ such that it corresponds to the average change in system volume due to the presence of such a bubble:
\begin{equation}\label{eq:vmethod1}
 v(\xi) = \frac{\partial}{\partial n} \langle V \rangle_{n(\xi)}.
\end{equation}
Here, $\langle V \rangle_{n(\xi)}$ is the average volume of the system when $n$ bubbles of volume $\xi$ are present and $v(\xi)$ corresponds to the average change in system volume $V$ when a single bubble of volume $\xi$ is added or removed. For bubble volumes that are unlikely to occur spontaneously on the time-scale of a straightforward simulation, $n(\xi)$ is either zero or one and there are no larger bubbles in the system such that Eq.~(\ref{eq:vmethod1}) becomes
\begin{equation}\label{eq:vmethod}
 v(\xi) = \langle V \rangle_\xi - \langle V \rangle.
\end{equation}
Here, $\langle V \rangle_\xi$ is the average volume of the system under the constraint that the largest bubble has a preliminary volume of $\xi$ and $\langle V \rangle$ is the average volume of the unconstrained metastable liquid. 
On the average, this definition of the bubble volume $v$ corresponds to choosing the equimolar dividing surface between the liquid and the (empty) interior of the bubble, which guarantees thermodynamic consistency. In particular, this definition of bubble volume fulfills the nucleation theorem \cite{KashchievJCP1982} (a proof is provided in the Appendix) and $pv$ corresponds to the average mechanical work of expanding the system due to the bubble. Note that, although the calibration procedure is based on the average volume of the system when a number of bubbles of a given size is present, the detection of the bubbles on the grid is still a local procedure and thus suitable for the study of large systems, where fluctuations in the liquid density may render global order parameters ineffective.
Further, due to the calibration procedure, the V-method is independent of arbitrary parameters associated with the detection of vapor-like grid-points, provided that the resolution of the three-dimensional grid is reasonably high. 
\begin{figure}[t]
\centering
 \includegraphics[width=0.4\textwidth]{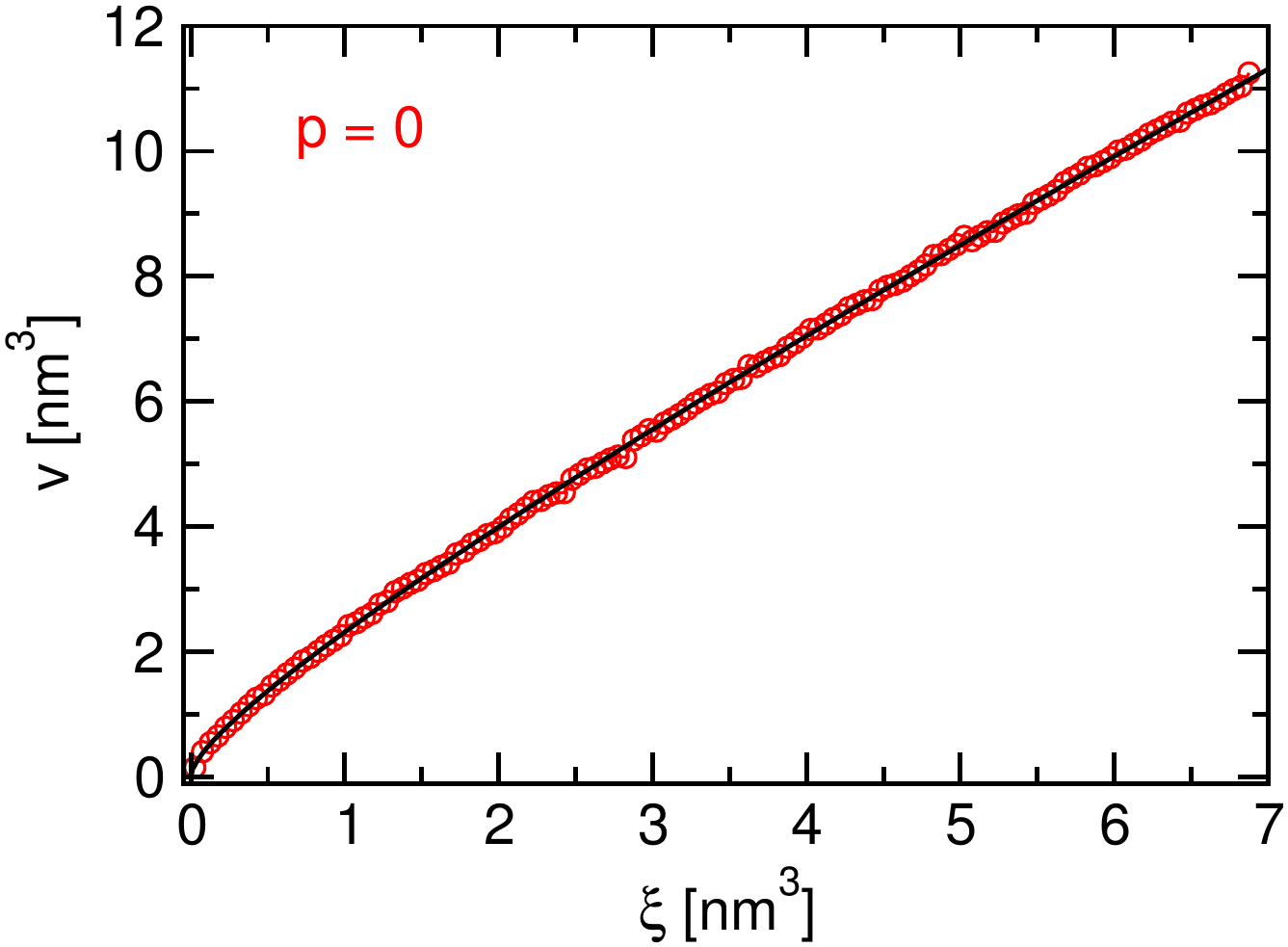}
 \caption{Estimate for the bubble volume $v$ (red circles) as a function the preliminary volume estimate $\xi$ determined using a grid-based approach. The fit (black line) according to Eq.~(\ref{eq:fit2}) maps $\xi$ onto the average change $v$ in system volume when such a bubble is present.}
 \label{fig:vofvp0}
\end{figure}
In practice, we fit $v(\xi)$ with a function of the form
\begin{equation}\label{eq:fit2}
 v(\xi) \approx \xi + k_1 \xi^{\frac{2}{3}} + k_2 \xi^{\frac{1}{3}}\, ,
\end{equation}
where the fit parameters $k_1$ and $k_2$ depend on the chosen thermodynamic state point. As shown in Ref.~\onlinecite{ournucleationpaper}, a single mapping from $\xi$ onto $v(\xi)$ with 
$k_1 \approx 1.04 \,{\rm nm}$ and $k_2 \approx 0.33 \,{\rm nm^2}$
is very accurate for all negative pressures studied here.
At zero pressure, we obtain an improved fit with the slightly altered parameters $k_1 \approx 1.02 \,{\rm nm}$ and $k_2 \approx 0.29 \,{\rm nm^2}$ (shown in Fig.~\ref{fig:vofvp0}).
\subsection{Entropy of the flat liquid--vapor interface}\label{sec:entropy}
As a reference point,
we compute the entropy of the flat liquid--vapor interface from the 
data for the surface tension $\gamma$ of TIP4P/2005 water at ambient pressure reported by Vega and de Miguel in Ref.~\onlinecite{VegaMiguelJCP2007}, where a heuristic functional form is fit to the simulation data. The derivative of $\gamma$ with respect to temperature yields an estimate for the surface entropy $s$ per unit area $A$:
\begin{equation}
 \frac{s}{A} = -\frac{\rm d \gamma}{{\rm d}T} = \frac{c_1}{9\, T_{\rm c}}\, \tau^{2/9} \left( 
  11 + 20\, c_2\, \tau 
 \right)  \, ,
\end{equation}
where $\tau = 1 - T/T_{\rm c}$, $c_1 = 227.86\, {\rm mJ/m^{-2}}$, $c_2 = -0.6413$, and the critical temperature $T_{\rm c} = 641.4\, {\rm K}$. For the studied temperature of $T = 296.4 \, {\rm K}$ this yields an estimate of $s / A= 10.28 \, k_{\rm B}/{\rm nm^2}$, where $k_{\rm B}$ is the Boltzmann constant.
\section{Results and discussion}
 
\begin{figure}[t]
\centering
 \includegraphics[width=0.4\textwidth]{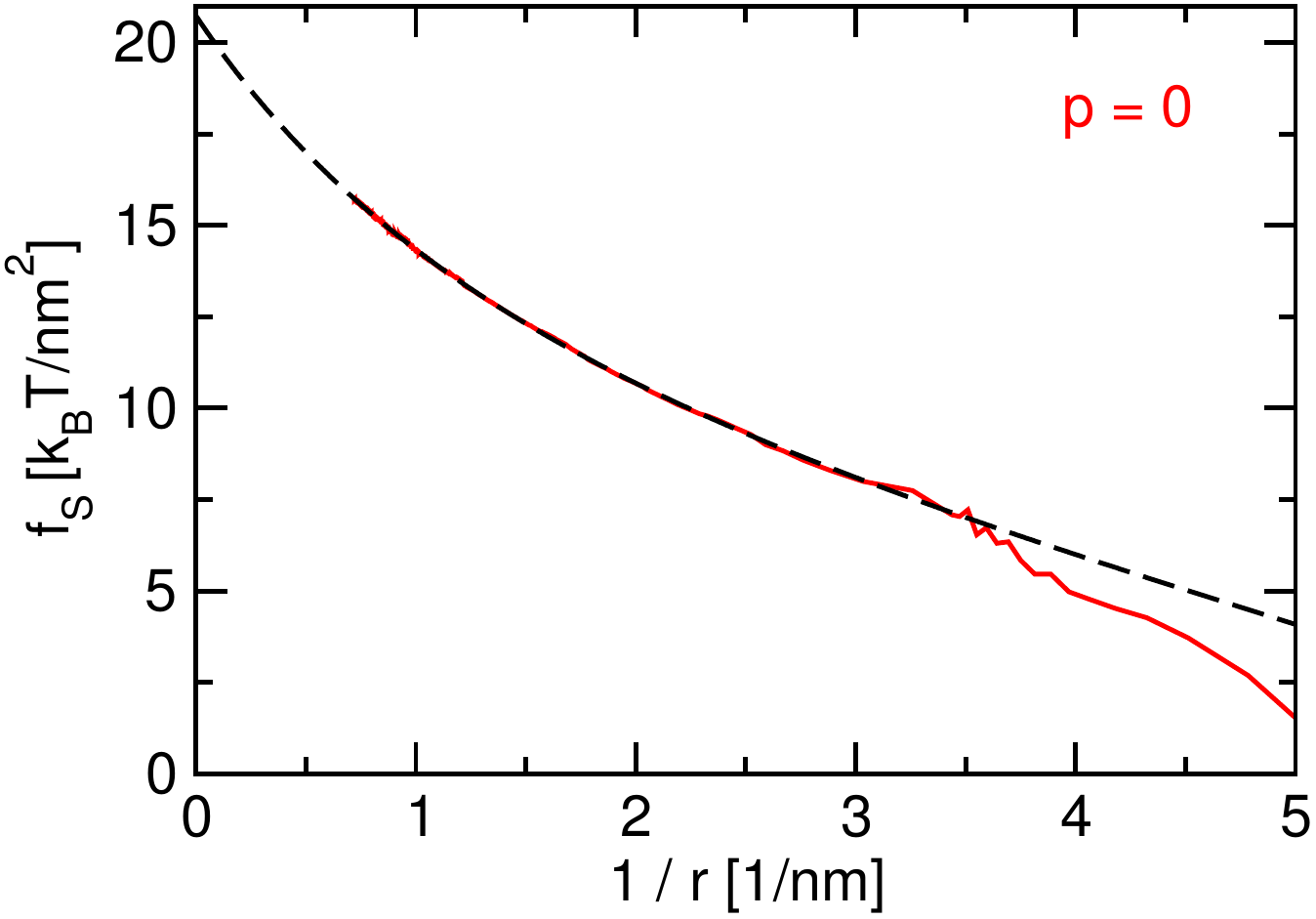}
 \caption{Surface free energy $F_{\rm S}$ as a function of inverse bubble radius $r^{-1}$ at zero pressure.
 The dashed black line is a fit to data in the range $0 < r^{-1} < 3.5\, {\rm nm^{-1}}$ using a Tolman-like form. }
 \label{fig:surfacefep0}
\end{figure}

\begin{figure}[t]
\centering
 \includegraphics[width=0.4\textwidth]{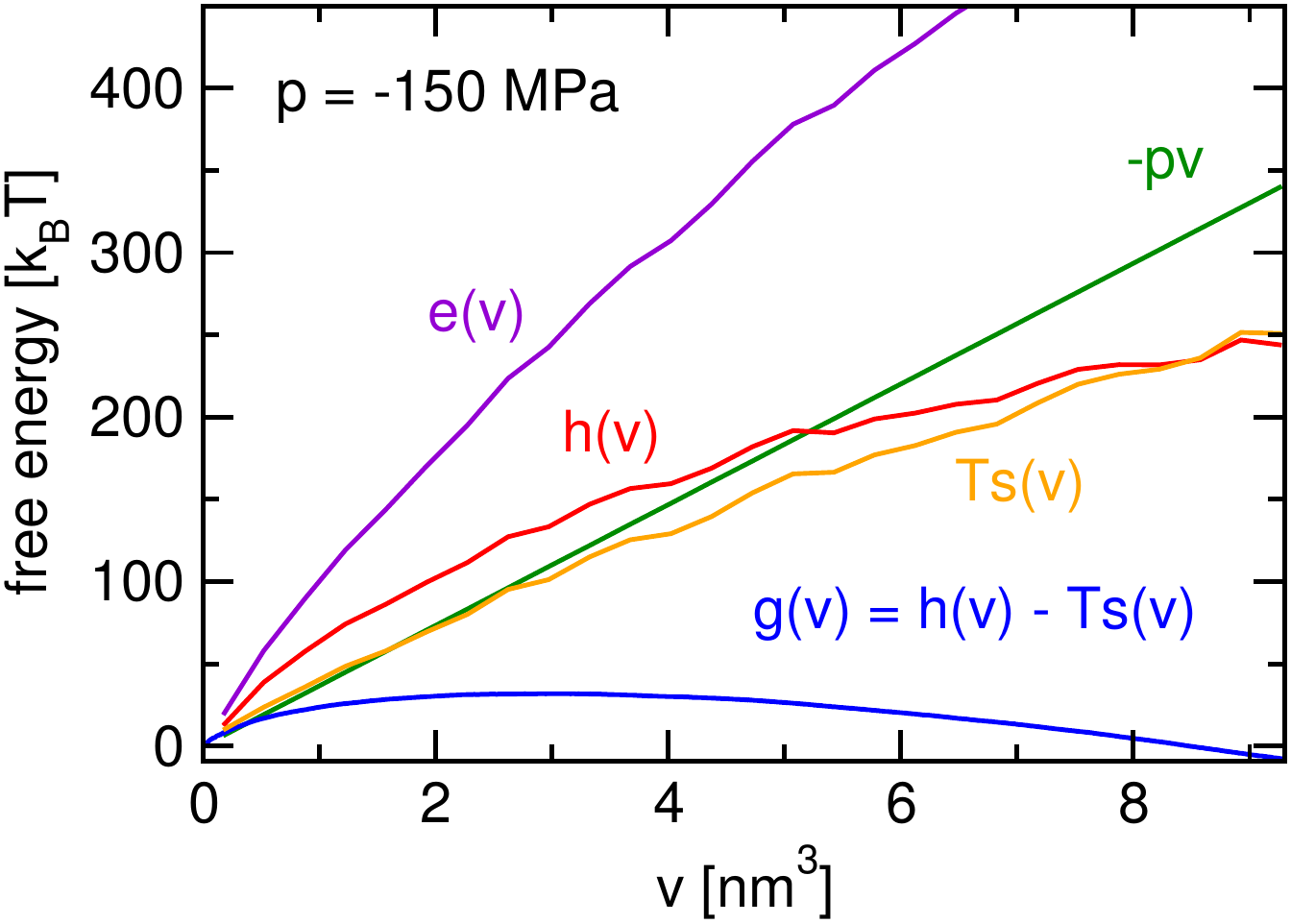}
 \caption{Comparison between the bubble free energy $g(v) = h(v) - T s(v)$ (blue), the enthalpy $h(v)$ (red), and the bubble entropy $Ts(v)$ (orange) as a function of bubble volume $v$ at $T = 296.4 \, {\rm K}$ and $p = -150 \, {\rm MPa}$. The enthalpy $h(v) = e(v) + pv$ is the sum of the internal energy $e(v)$ (purple), which increases with $v$ due to the energetic cost of the vapor--liquid interface, and the mechanical work $pv$ (green), which favors the formation of bubbles. The bubble free energy has a maximum at $v^\ast = 2.95 \, {\rm nm^3}$ while the enthalpy and the entropy both increase monotonically over the range of bubble volumes shown. Enthalpy and entropy are of comparable magnitude which illustrates the importance of entropy for cavitation at these conditions.}
 \label{fig:hvsg}
\end{figure}

\begin{figure}[t]
\centering
 \includegraphics[width=0.4\textwidth]{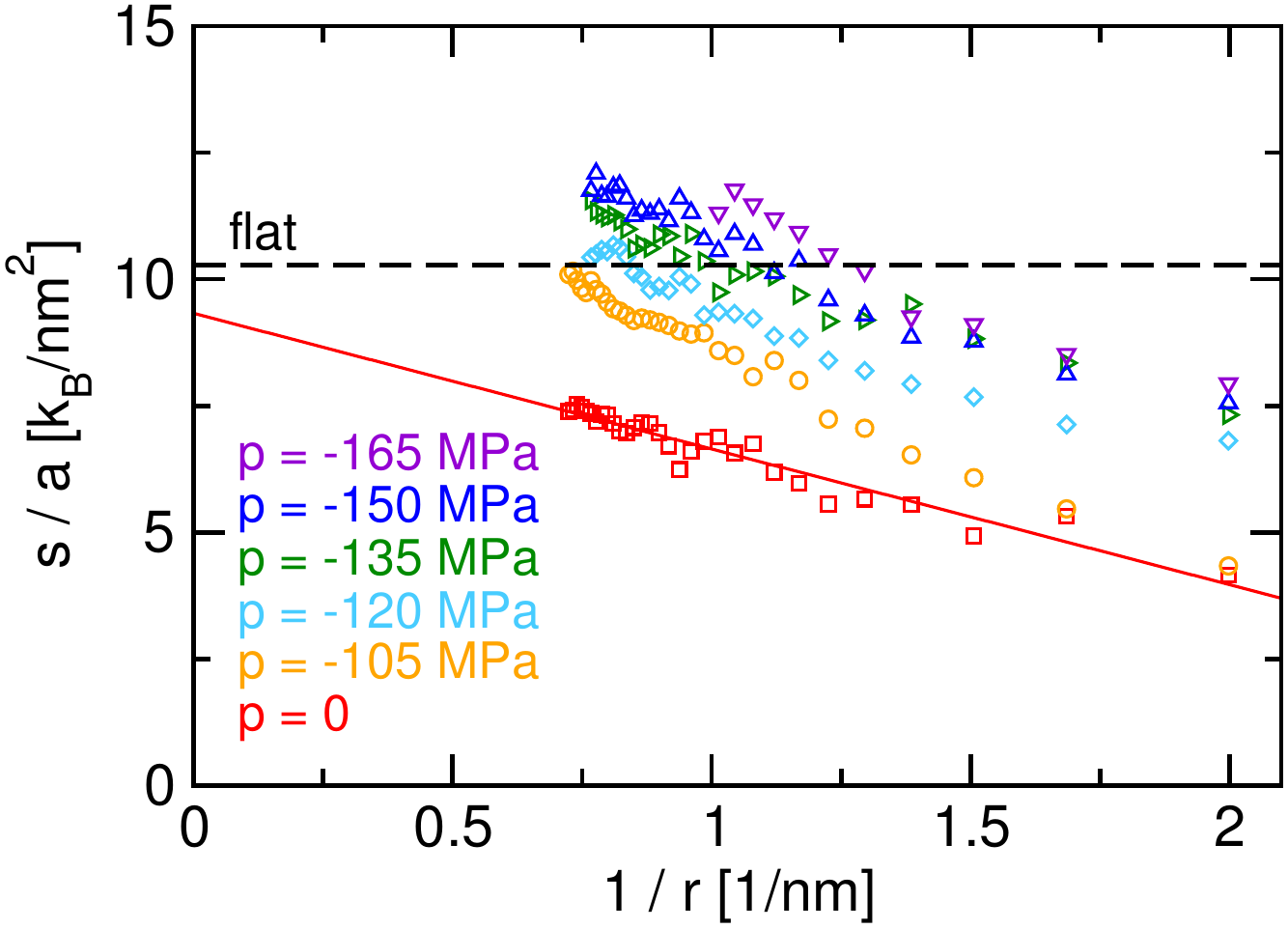}
 \caption{Bubble entropy per unit area $s/a$ as a function of mean curvature $1/r$ at a temperature $T = 296.4 \, {\rm K}$ and various tensions.
 Here, $r = [3 v / (4 \pi )]^{1/3}$ and $a = (36 \pi v^2)^{1/3}$ are the radius and surface of a sphere with volume $v$, respectively.
 The estimates for the entropy were obtained via Eq.~(\ref{eq:entropy}) by averaging the data from equilibrium umbrella sampling simulations over bins with a width of $0.35 \, {\rm nm^3}$. The black dashed line indicates the surface entropy of a planar liquid--vapor interface for the TIP4P/2005 water model \cite{VegaMiguelJCP2007} at ambient pressure and the red line is a linear fit to the data at zero pressure.}
 \label{fig:sofvofr}
\end{figure}

We obtain the entropy $s(v)$ associated with a bubble of size $v$ by computing the equilibrium free energy of bubble nucleation, $g(v)$, from umbrella sampling calculations and
subtracting all contributing terms other than the entropy:
\begin{equation}\label{eq:entropy}
 Ts(v) = -g(v) + e(v) + pv.
\end{equation}
Here, the bubble energy $e(v) = \langle E \rangle_v - \langle E \rangle$ is the difference in average energy of the system containing a bubble of size $v$, $\langle E \rangle_v$, and the unconstrained average in the metastable liquid, $\langle E \rangle$. 
The mechanical work gained by expanding the system at negative pressure is given by $pv$, since $v$ is calibrated to correspond to the average change in system volume due to the presence of a bubble (see Appendix).
The obtained estimate for the bubble entropy $s(v)$ depends on the chosen normalization of the free energy.
In order to obtain a normalization for the free energy $g(v)$ of nucleation, which also determines the normalization of the bubble entropy $s(v)$,
we examine the bubble surface free energy,
\begin{equation}
f_S = \frac{1}{a} (-k_{\rm B}T \ln [ v_0^2 \rho(v) ] - pv). 
\end{equation}
Here, $a = (36 \pi v^2)^{1/3}$ is the surface area of a sphere with volume $v$, $v_0 = 1 \, {\rm nm^3}$ determines the unit of volume, and $\rho(v)$
is the density of bubbles with volume $v$. The bubble density $\rho(v)$ is defined such that $\rho(v){\rm d}v$ is the average number of bubbles with volumes in the interval $[v, v+{\rm d}v]$ divided by the total volume $V$ of the system.
Note that $-k_{\rm B}T \ln [ v_0^2 \rho(v) ]$ is equal to the free energy of cavitation $g(v)$ up to a normalization constant that will be determined in the following.
By subtracting the mechanical work $pv$, which drives the transition from the liquid to the vapor, the resulting surface free energy is simply the free energetic cost of the liquid--vapor interface per surface area.
The surface free energy at zero pressure is shown in Fig.~\ref{fig:surfacefep0} as a function of $1/r$, where $r = [3v/(4 \pi)]^{1/3}$ is the radius of a sphere with volume $v$ (surface free energies for negative pressures have been obtained previously in Ref.~\onlinecite{ournucleationpaper}).
The surface free energy can be fitted by a function that takes the curvature dependence of the surface tension $\gamma(r)$ into account, akin to the one proposed by Tolman \cite{Tolman},
\begin{equation}
\label{eq:fefit}
f_S(r) = \frac{\gamma_0}{1 + 2 \delta / r} + \frac{C}{4 \pi r^2}.
\end{equation}
At zero pressure, the fit yields values of $\gamma_0 = 20.73 \, k_{\rm B} T /{\rm nm^{2}}$, $\delta = 0.218 \, {\rm nm}$, and $C = -0.0972 \, k_{\rm B} T$ (the fit is indicated by the dashed line in Fig.~\ref{fig:surfacefep0}).
The surface free energies over the range of negative pressures studied here look qualitatively similar.
Furthermore, as shown in Ref.~\onlinecite{ournucleationpaper}, 
the surface free energy at all negative pressures studied here can be fit by a single curve, where the obtained parameters were $\gamma_0 =  20.24 \, k_{\rm B} T/{\rm nm^{2}}$ and $\delta = 0.195 \, {\rm nm}$, which are both similar to but slightly lower than the values obtained at zero pressure, and $C =  -3.80\, k_{\rm B} T$.
We will use Eq.~(\ref{eq:fefit}) to normalize the cavitation free energy $g(v)$ such that it goes to zero in the limit of vanishing bubble size, i.e., $\lim_{v \rightarrow 0} g(v) = 0$, by choosing $g(v) = -k_{\rm B}T \ln [ v_0^2 \rho(v) ] - C$.
\footnote{Alternatively, one could in principle fit the surface free energy for very small bubbles with a suitable function and extrapolate this fit to zero to obtain the normalization of the free energy.
However, although the order parameter employed here is calibrated such that it yields a physically meaningful estimate for the volume of a bubble, the {\it smallest} cavity it can detect depends on the chosen parameters employed in the grid-based detection of bubbles in the system (namely, the spatial resolution of the grid and the chosen distance within which a grid-point is considered to be occupied by nearby water molecules). 
As such, if one were to simply fit the free energy for low $v$ and extrapolate this fit to zero, the obtained estimate for the free energy of cavitation would depend on the arbitrary parameters employed in the detection of bubbles in the system.
Since the obtained normalization affects the height $g(v^\ast)$ of the free energy barrier, which is connected to the probability of encountering a bubble of critical size (and thus to the cavitation rate), this dependence on the detection of minute cavities is clearly unphysical.
Consequently, in order to mitigate the impact of arbitrary parameters on the estimate for the free energy of cavitation, we extrapolate the fit given in Eq.~(\ref{eq:fefit}), which reproduces the free energy very accurately over a wide range of bubble volumes, to zero to obtain the normalization of the free energy.}
This choice of normalization is equivalent to the notion that the formation of a vanishingly small bubble does not carry a free energetic cost.
Since the other terms in Eq.~(\ref{eq:entropy}) contributing to the bubble entropy $s(v)$, namely the average internal energy $e(v)$ and the mechanical work $pv$ associated with a largest bubble of size $v$, fulfill this criterion by construction, this choice implies $\lim_{v \rightarrow 0} s(v) = 0$. As such, the bubble entropy presented here does not contain the translational entropy, corresponding to realizing a set of identical configurations where all molecules (and thus the bubble) have been shifted, or effects from the width of the histograms used to compute the free energy in umbrella sampling, thereby making it compatible with the role of bubble entropy in classical nucleation theory.
At finite temperatures, the limit of metastability of a liquid under tension is determined by the competition of enthalpy, which favors the metastable liquid, and entropic contributions, which drive the system towards the vapor phase. 
The relative importance of the enthalpic and entropic contributions to the free energy barrier impeding bubble formation depends on the state-point. For very low temperatures, when the influence of entropy is insignificant, the limit of metastability is reached when the tension acting upon the system is high enough that the gain in mechanical work $pv$, where $v$ is the bubble volume, compensates for the increase in energy $e(v)$ due to this bubble. 
At ambient temperatures however, the entropy gained from forming a bubble has a significant impact on the limit of metastability, 
and at a tension of $p = -150 \, {\rm MPa}$, the free energy $g(v)$ exhibits a maximum at $\sim 3 \, {\rm nm^3}$ as illustrated in Fig.~\ref{fig:hvsg}.
Bubbles larger than this critical volume typically grow and the system transitions to the vapor phase. In contrast, the enthalpy rises steadily with increasing bubble volume and its maximum lies significantly beyond the investigated range of bubble volumes, with a barrier orders of magnitude larger than the one actually present when entropic contributions are taken into account. As such, bubble entropy is the driving force behind cavitation in water over a wide range of conditions relevant in biology
\cite{Holbrook2012,StroockHolbrookRevFluidMech2014,PonomarenkoMarmottantRoyalSoc2014,LarterDelzonPlantPhys2015,RowlandOliveiraNature2015,WheelerStroockNature2008,VincentOhlPRL2012,OlivierOhlSoftmatter2014,snappingshrimp,PatekCaldwell2005,IosilevskiiWeihs2008,NoblinDumaisScience2012,OhlLohseBiophysJ2006,AdhikariBerkowitzJPhysChemB2015}
and engineering \cite{Trevena1987,Brennen1995,FrancMichel,Kumar2010}.
Since the bubbles are essentially completely empty cavities in the liquid at the conditions studied here, their entropy is exclusively a property of the liquid--vapor interface.
The bubble entropy per unit area, $s/a$, is shown in Fig.~\ref{fig:sofvofr} as a function of $1/r$.
Over the studied range of tensions, the bubble entropy per unit area decreases approximately linearly with the mean curvature $1/r$ of the bubble.
As a point of comparison, an estimate for the entropy of a planar interface at ambient pressure (see Sec.~\ref{sec:entropy}), whose value exceeds the 
linear extrapolation of $s$ vs. $1/r$ at zero pressure to vanishing curvature by about $1\,k_{\rm B}$, is indicated by the dashed line in Fig.~\ref{fig:sofvofr}.
The bubble entropy $s$ is positive, thus favoring the formation of bubbles. While this is true in general for transitions from the liquid to the vapor, it is worth keeping in mind that the entropic gain from vapor-like molecules in the bubble is insignificant.
Furthermore, since the definition of the bubble entropy given above removes contributions from the translational entropy, 
there are two likely sources for the pronounced entropic gain from bubble formation: fluctuations in the bubble shape and the increased rotational entropy of interfacial OH-groups when compared to the bulk.
Interestingly, the number of free OH-groups, i.e., OH-groups that do not partake in a hydrogen bond, per molecule at the interface shows the same curvature dependence as the bubble entropy per unit area (data shown in Ref.~\onlinecite{ournucleationpaper}).
Assuming that the bubble entropy is governed by the entropic gain due to increased rotational freedom of interfacial OH-groups, the bubble entropy per free OH group evaluates to about $2.5 - 3.5 \, k_{\rm B}$ for all bubble sizes depending on pressure, comparable to the entropic gain $\sim 4 \, k_{\rm B}$ estimated for a dangling OH-group at the end of a single-file chain of water molecules.\cite{KoefingerDellagoJCP2009}
The estimates for the entropy of bubble formation obtained here provide the basis for the use of the nucleation theorem in the interpretation of experimental data obtained by cooling at constant volume.
When the nucleation theorem is invoked to estimate the volume $v^\ast$ of the critical bubble from cavitation rate measurements,\cite{AzouziCaupinNaturePhys2013}
one first assumes that the cavitation rate $J$, i.e., the number of cavitation events per volume and time, has the form $J = \nu \exp[-\beta g(v^\ast)]/(V \Delta t)$. Here, a measurement is performed on a system of volume $V$ over a time $\Delta t$ and $\nu$ is a kinetic pre-factor that accounts for the dynamics of the transition.
The choice of this kinetic pre-factor, which relates the rates measured at various temperatures to the height $g^\ast$ of the free energy barrier, is somewhat arbitrary and various choices are described in the literature \cite{BlanderKatzAIChE1975,PettersenMarisPRB1994}; since the height $g(v^\ast)$ of the free energy barrier enters the rate estimate exponentially and the prefactor is supposed to vary only weakly with temperature, it is usually assumed that the observed change in the cavitation rates is governed by changes in $g(v^\ast)$, diminishing the importance of the exact choice of $\nu$.
To obtain $v^\ast$ by use of the nucleation theorem,\cite{KashchievJCP2006}
\begin{equation}\label{eq:nuctheorem_exp}
v^\ast = \left( \frac{\partial p}{\partial T} \right)_V^{-1} \left( \frac{{\rm d} g^\ast}{{\rm d} T} + s^\ast \right) ,
\end{equation}
one needs to estimate ${\rm d} g^\ast/{\rm d} T$ from cavitation rates measured at different temperatures.
In addition, one requires the quantities $\left( \partial p / \partial T \right)_V$ and $s^\ast$, where the first term can be estimated by extrapolating an equation of state to the conditions of interest.\cite{CaupinHerbertCRPhys2006,DavittBalibarJCP2010,PallaresCaupinPCCP2016}
This leaves the entropy $s^\ast$ of the critical bubble that we computed for various pressures at a temperature $T = 296.4 \, {\rm K}$. However, if one aims to analyze data from experiments conducted along an isochoric path, the entropy over a range of temperatures is required.
The entropy of a flat liquid--vapor interface at ambient pressure can be employed as an approximation for the entropy $s^\ast$ of the critical bubble to determine its volume $v^\ast$ from experimental data obtained along an isochoric path via Eq.~(\ref{eq:nuctheorem_exp}).
Since the bubble entropy decreases roughly with $1/r$, taking its value for a flat interface {\em overestimates} its true value for a bubble of finite size (see Fig.~\ref{fig:sofvofr}). However, as the bubble entropy increases with tension, using bubble entropies obtained at ambient pressure would in turn {\em underestimate} its magnitude.
As a result, the entropy $-{\rm d}\gamma/{\rm d} T$ of a flat interface at ambient pressure (indicated by the black dashed line in the figure, see Sec.~\ref{sec:entropy}), where data on the temperature dependence of the surface tension is readily available, appears to be a reasonable estimate for the entropy of bubbles in water under tension over the range of critical bubble volumes $v^\ast$ which are physically relevant, i.e., at conditions where rate measurements can be obtained on time-scales accessible to experiments \cite{ournucleationpaper}.
In order to assess the accuracy of this approach, we mimick the procedure employed to estimate $v^\ast$ from experimental data utilizing this approximation and compare the obtained estimates to the value of $v^\ast$ measured directly in simulations, i.e., the location of the maximum of $g(v)$ at different pressures.
To this end, we first compute $(\partial p/\partial T)_V = -(\partial V/\partial T)_p / (\partial V/\partial p)_T$ required in Eq.~(\ref{eq:nuctheorem_exp}). Here, $(\partial V/\partial T)_p$ is obtained by measuring the average volume $\langle V \rangle$ of the metastable liquid at various temperatures in the range $290.4$--$302.4 \, {\rm K}$ and taking the slope of a linear fit to the data at $T = 296.4 \, {\rm K}$. Similarly, we obtain an estimate for $(\partial V/\partial p)_T$ by fitting $\langle V \rangle$ as a function of pressure with a polynomial function and taking the pressure derivative of the fit.
We combine the direct measurement of $v^\ast$ from simulation, $(\partial p/\partial T)_V$, and $s^\ast$ (from the data presented in Fig.~\ref{fig:sofvofr}) to obtain an estimate for ${\rm d} g^\ast/{\rm d} T$ from Eq.~(\ref{eq:nuctheorem_exp}).
We now have all that is needed to compute an estimate for $v^\ast$ determined by solving
\begin{equation}\label{eq:nuctheorem_approx}
 v^\ast \left(\frac{\partial p}{\partial T}\right)_V + v^{\ast \frac{2}{3}} (36 \pi)^\frac{1}{3} \left(\frac{\partial \gamma}{\partial T}\right) - \frac{{\rm d}g^\ast}{{\rm d}T} = 0.
\end{equation}
(This equation is just Eq.~(\ref{eq:nuctheorem_exp}) with the approximation $s^\ast = - v^{2/3}(36 \pi)^{1/3} \partial \gamma / \partial T$, where $\partial \gamma/\partial T$ is evaluated for a flat liquid--vapor interface at ambient pressure.)
\begin{figure}[t]
\centering
 \includegraphics[width=0.4\textwidth]{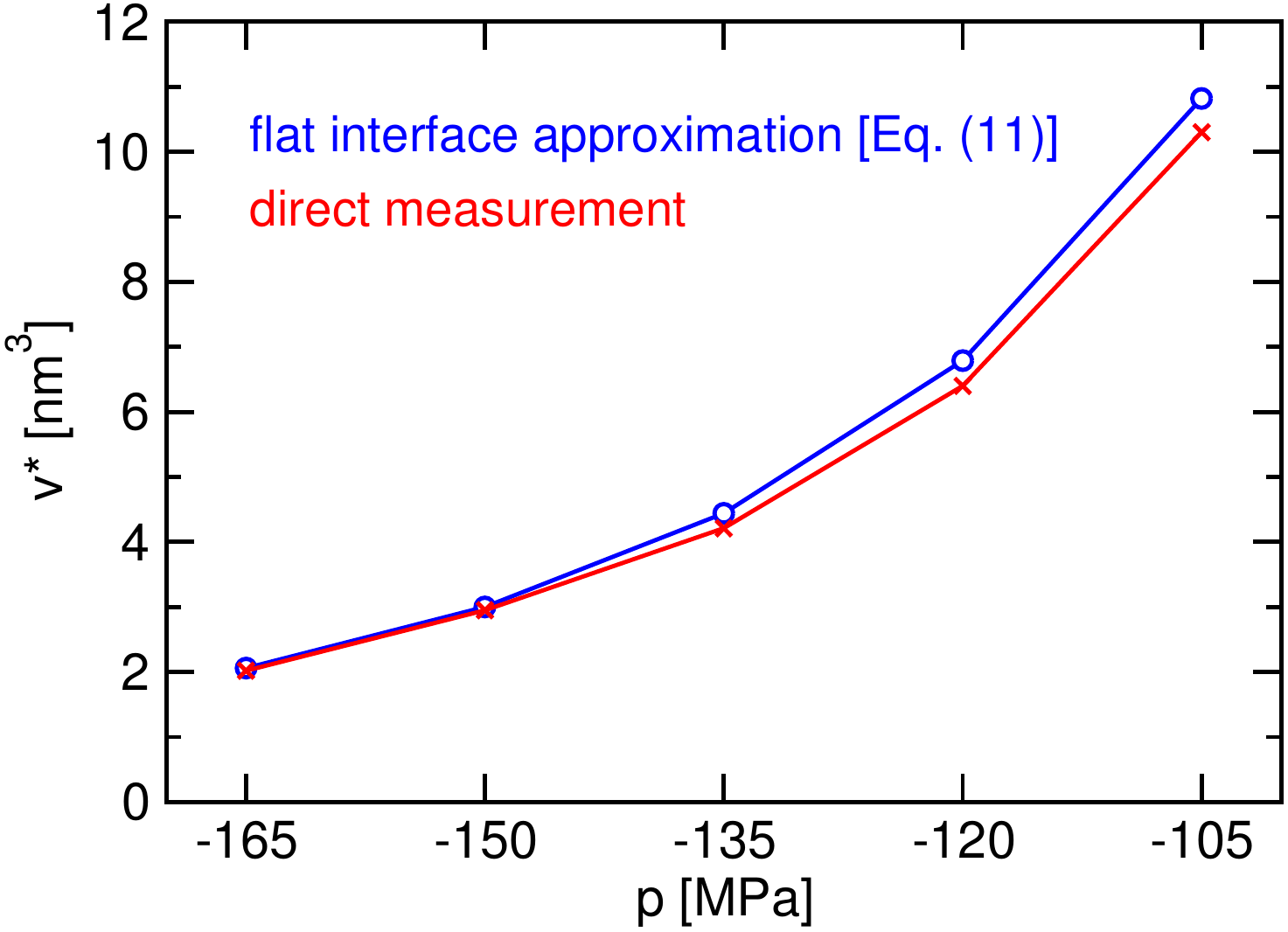}
 \caption{Estimates for the volume $v^\ast$ of the critical bubble obtained using Eq.~(\ref{eq:nuctheorem_approx}) and directly from simulation, respectively. The estimate computed using the entropy $ - \partial \gamma / \partial T$ of the flat liquid--vapor interface at ambient pressure as an approximation (blue circles) agrees well with that directly determined from simulation (red crosses, data from Ref.~\citenum{ournucleationpaper}).}
 \label{fig:vstarcomp}
\end{figure}
The resulting estimates for $v^\ast$ are shown in Fig.~\ref{fig:vstarcomp} alongside those obtained directly from the free energy profile determined by simulation.
Over the studied range of tensions and at ambient temperature, approximating the entropy $s^\ast$ of the critical bubble using the entropy per unit area of the flat interface at ambient pressure leads to accurate predictions for the volume $v^\ast$ of the critical bubble.
Although the approximation overestimates $v^\ast$ consistently, the deviation is small (the relative error ranges from $2$ to $6\%$) and the approximation becomes more accurate for stronger tensions.
Assuming $s$ behaves similarly in the temperature range 
$322$--$335 \, {\rm K}$ 
investigated in Ref.~\onlinecite{AzouziCaupinNaturePhys2013}, we are now in a position to obtain an improved estimate for the volume $v^\ast$ of the critical bubble.
Choosing $T = 328\, {\rm K}$, in the middle of the investigated range of temperatures, and solving Eq.~(\ref{eq:nuctheorem_approx}) for $v^\ast$, we obtain $v^\ast = 4.74 \, {\rm nm^3}$ for the volume of the critical bubble.
This estimate was calculated using
the quantities $(\partial p/ \partial T )_V = 61.45\, k_{\rm B} / {\rm nm^2}$, ${\rm d}g^\ast / {\rm d}T = 0.4116 \, k_{\rm B}T / {\rm K}$ (both values taken from Ref.~\onlinecite{AzouziCaupinNaturePhys2013}), and $ \partial \gamma / \partial T = -11.54\, k_{\rm B} / {\rm nm^2}$ (from Sec.~\ref{sec:entropy}).
\section{Conclusions}
At ambient temperature, the entropic gain associated with the formation of a bubble has a significant influence on the free energy of cavitation.
The entropy of bubble formation is positive and over the investigated range of bubble volumes its magnitude is comparable to that of the enthalpy, thereby significantly reducing the metastability of water under tension.
For the tensions investigated here, the entropy of a bubble decreases approximately linearly with its average curvature and the entropic gain due to the formation of a bubble is more pronounced at higher tensions.
For critical bubble volumes relevant in cavitation experiments, the entropy of a flat liquid--vapor interface at ambient pressure is a good approximation for the entropy of the critical bubbles. 
Using this approximation to evaluate the experimental data reported in Ref.~\onlinecite{AzouziCaupinNaturePhys2013} yields
a critical bubble volume of $v^\ast \approx 5\, {\rm nm^3}$,
confirming the estimate reported therein.
\section{Acknowledgements}
This work was supported by the Austrian Science Fund (FWF) under grant P24681-N20 and SFB Vicom F41.
The authors thank J.~L.~F.~Abascal, F.~Caupin, P.~Geiger, M.~A.~Gonzalez, and C.~Valeriani for insightful comments.
The computational results presented have been achieved using the Vienna Scientific Cluster (VSC).
\appendix*
\section{Nucleation theorem and bubble volume}
\label{sec:nuctheorem}
The nucleation theorem at constant temperature relates the change in the barrier height $g^\ast = g[v^\ast(p)]$ with pressure to the volume $v^\ast(p)$ of the critical bubble,
\begin{align}\label{eq:app_nuctheoremconstT}
  \frac{\mathrm{d} g^\ast}{\mathrm{d} p} &= \frac{\partial g}{\partial v} \bigg\vert_{v^\ast} \frac{\partial v^\ast}{\partial p} + \frac{\partial g}{\partial p}\bigg\vert_{v^\ast} \nonumber \\&= v^\ast.
\end{align}
Here, the first term on the right hand side of the first line vanishes since the free energy $g(v)$ has a maximum at the volume $v^\ast$ of the critical bubble by definition, leaving only the partial derivative $\partial g(v)/\partial p\rvert_{v^\ast} = v^\ast.$ 
In the following, we provide an analytical expression for $\partial g(v)/\partial p\rvert_{v^\ast}$ and show that the 
definition of the bubble 
volume $v$ as defined in Eq.~(\ref{eq:vmethod}) is compatible with the nucleation theorem.
The free energy of cavitation is determined as a function of $\xi$, the volume estimate obtained directly from a grid-based \cite{WangFrenkelJPhysChemB2009} approach:
\begin{equation}\label{eq:g_app}
 g(\xi) = - k_{\rm B} T \ln [P(\xi)].
\end{equation}
Here,
\begin{align}
 P(\xi) &= \frac{1}{Q} \int {\rm d}x {\rm d} V \, e^{-\beta [E(x) + pV]}\delta [ \xi(x) - \xi ] \nonumber \\
 &= \langle \delta [ \xi(x) - \xi ] \rangle
\end{align}
is the probability density that the largest bubble in the configuration $x$ has a volume of $\xi$, where $\delta$ is the Dirac delta function and 
$Q = \int {\rm d}x {\rm d} V \, e^{-\beta [E(x) + pV]}$ is the partition function. The angular brackets $\langle \cdot \rangle$ denote an average in the {\it NpT }ensemble.
The change of $g(\xi)$ upon a change in pressure is given by
\begin{equation}\label{eq:app_dgdp}
 \frac{\partial g(\xi)}{\partial p} = - \frac{k_{\rm B} T}{P ( \xi )} \frac{\partial P(\xi)}{\partial p}.
\end{equation}
Carrying out the derivative with respect to pressure yields
\begin{align}
  \frac{\partial P(\xi)}{\partial p} &= 
 -\beta \left[ \langle V \delta [ \xi(x) - \xi ] \rangle - \langle V \rangle \langle \delta [ \xi(x) - \xi ] \rangle \right].
\end{align}
Inserting this into Eq.~(\ref{eq:app_dgdp}) leads to
\begin{align}
 \frac{\partial g(\xi)}{\partial p} &= \frac{\langle V \delta [ \xi(x) - \xi ] \rangle}{\langle \delta [ \xi(x) - \xi ] \rangle} - \langle V \rangle \frac{\langle \delta [ \xi(x) - \xi ] \rangle}{\langle \delta [ \xi(x) - \xi ] \rangle} \nonumber \\
 &= \langle V \rangle_\xi - \langle V \rangle = v(\xi),
\end{align}
where the notation $\langle \cdot \rangle_\xi$ indicates an average under the constraint $\delta [ \xi(x) - \xi ]$ and $v$ is the estimate for the volume of the largest bubble in the system according to Eq.~(\ref{eq:vmethod}).
Since Eq.~(\ref{eq:fit2}) uniquely maps each value of $\xi$ onto a value $v$, the constrained averages $\langle \cdot \rangle_\xi$ and $\langle \cdot \rangle_v$ are identical and by transforming coordinates 
one obtains
\begin{align}
 \left. \frac{\partial g(v)}{\partial p} \right\rvert_{v^\ast} = v^\ast,
\end{align}
where $g(v) =  g(\xi) + k_{\rm B}T \ln(\vert {\rm d}v / {\rm d} \xi \vert)$. Thus, the definition of the bubble volume $v$ employed here is consistent with the nucleation theorem.
For cavitation along an isochoric path, where the tension on the liquid is varied by changing the temperature at constant volume, the nucleation theorem states
\begin{align}
  \frac{\mathrm{d} g^\ast}{\mathrm{d} T} &= 
  \frac{\partial g}{\partial v} \bigg\vert_{v^\ast} \frac{\partial v^\ast}{\partial T} + \frac{\partial g}{\partial p}\bigg\vert_{v^\ast} \left(\frac{\partial p}{\partial T}\right)_V + \frac{\partial g}{\partial T}\bigg\vert_{v^\ast}\nonumber \\
  &= v^\ast \left(\frac{\partial p}{\partial T}\right)_V - s^\ast,
\end{align}
where $s^\ast = s(v^\ast)$ is the entropy of the critical bubble.
Here, like in Eq.~(\ref{eq:app_nuctheoremconstT}), the first term on the right hand side of the first line vanishes since $g(v)$ has a maximum at $v^\ast$.
Thus, in the isochoric case, the nucleation theorem relates the temperature derivative of the barrier height to the volume and the entropy of the critical bubble.
\end{document}